\documentclass[runningheads, 12pt]{llncs} % Springer LNCS format
\usepackage[a4paper,margin=1in]{geometry} % Adjust margins to fit more words
\usepackage{amsfonts,amsmath,amssymb,bm}
\usepackage{graphicx}
\usepackage{siunitx}
\usepackage{hyperref}
\usepackage{booktabs}
\usepackage{float}
\usepackage{lscape}
\usepackage[T1]{fontenc}
\usepackage[default]{lato}

\title{Response Rate Estimation in Single-Stage Basket Trials: \\ A Comparison of Estimators That Allow for Borrowing Across Cohorts}
\titlerunning{Response Rate Estimation in Basket Trials} % Short title

\author{Antonios Daletzakis\inst{1,3} \and 
	Rutger van den Bor\inst{2} \and 
	Vincent van der Noort\inst{1} \and 
	Kit CB Roes\inst{3}}

\authorrunning{Daletzakis et al.} % Short author list

\institute{
	Biometrics Department, Netherlands Cancer Institute, Amsterdam, Netherlands \\ 
	\email{a.daletzakis@nki.nl} % Corresponding author
	\and
	Dept. of Data Science and Biostatistics, Julius Center for Health Sciences and Primary Care, University Medical Center Utrecht, Utrecht, Netherlands
	\and
	Department of Health Evidence, Section Biostatistics, Radboud University Medical Center, Nijmegen, Netherlands
}

\begin{document}
	
	\maketitle % Springer LNCS requires \maketitle after \title and \author
	
	\begin{abstract}
		Therapeutic advancements in oncology have transitioned towards targeted therapy based on specific genomic aberrations. This shift necessitates innovative statistical approaches in clinical trials, notably in the emerging paradigm of master protocol studies. Basket trials, a type of master protocol, evaluate a single treatment across cohorts sharing a common genomic aberration but differing in tumor histology.	While offering operational advantages, the analysis of basket trials introduces challenges with respect to statistical inference. Basket trials can be used to decide for which tumor histology the target treatment is promising enough to move to confirmatory clinical evaluation and can employ a Bayesian design to support this decision making. In addition to decision making, estimation of the cohort-specific response rates is highly relevant to inform the design of subsequent trials. This study evaluates seven Bayesian estimation methods for basket trials with a binary outcome, contrasted with the (frequentist) sample proportion estimate, through a simulation study. The objective is to estimate cohort-specific response rates, with a focus on average bias, average mean squared error, and the degree of information borrowing. A variety of scenarios are explored, covering homogeneous, heterogeneous, and clustered response rates across cohorts. The performance of the evaluated methods shows considerable trade-offs in bias and precision, emphasizing the importance of method selection based on trial characteristics. Berry's method excels in scenarios with limited heterogeneity. No clear winner emerges in a more general scenario, with method performance influenced by the amount of shrinkage towards the overall mean, bias, and the choice of priors and tuning parameters in more complex settings. Challenges include the computational complexity of methods, the need for careful tuning of parameters and prior distribution specification, and the absence of clear guidance on their selection. Researchers should consider these factors in designing and analyzing basket trials.
	\end{abstract}
	
	 % Springer LNCS uses \keywords

\section{Introduction}
% Begining of the introduction
%%%%%%%%%%%%%%%%%%%%%%%%%%%%%%%%%%%%%%%%
	Therapeutic approaches in oncology have shifted from conventional chemotherapy to targeted therapy on a specific genomic aberration across different cancer types. The progress in the field of precision medicine required a different statistical approach to clinical trials. Master protocol studies appear to become increasingly common in practice, particularly in non-randomized exploratory phase I/II research. \cite{Meyer} \cite{Park} An example of a master protocol study is the basket trial design, the focus of this paper. In such trials, the same treatment is evaluated in multiple cohorts of patients with different tumour histologies that share the same genomic aberration.
 
    Basket trials can be seen as a series of single-arm trials (usually designed as either single or two-stage) performed in distinct histology-based patient cohorts. An operational advantage of this design compared to conducting multiple individual trials for each tumor histology, is that it requires only a single protocol, a single database, and a single medical ethical review board application. As such it results in time and cost savings. With respect to the analysis of studies with a basket trial design, a straightforward inference strategy is to analyse each cohort separately, or to pool the data and provide a total estimate for the patients regardless of the patient's tumour histology. Either pooling or independently analyzing trial results in practical applications may not necessarily represent the optimal choice in all cases. When there is high heterogeneity the pooling estimate introduces type I error inflation and in low heterogeneity the independent analysis lacks of statistical power compared to alternative methods. A third option is to use a procedure which, like independent analyses, performs inference on individual cohorts, but which allows for borrowing information across cohorts. Such methods may provide advantages in terms of statistical efficiency. (see, e.g. Pohl et al. (2021)\cite{Pohl} for a comprehensive review). 
	
	 Methods or designs that allow for borrowing information are typically based on Bayesian procedures and assume that the primary outcome of interest is the response rate: the fraction of patients in each cohort who showed, upon treatment, a clinically significant shrinkage in their tumor volume, a common endpoint in early phase oncology trials. Although the primary focus of the methods review in current article is on decision-making, posterior point estimates of the cohort-specific response rates can be derived as well. Such estimates are of relevance to plan subsequent trials or to provide estimates to support benefit-risk assessment and communicate expected effects to patients. While operating characteristics of the decision-making process are important as well, the focus of the current simulation study is on the performance (bias, MSE) of the response rate estimator derived from the posterior distribution used in the Bayesian procedures.

     An overview of the performance of estimators based on seven Bayesian analysis methods that allow for borrowing is presented for a range of scenarios. Additionally, we investigate the influence of prior distribution choice on the performance of Bayesian estimators. All the estimators are applied in a variety of scenarios all considering parallel single stage cohorts. A range of scenarios is addressed, encompassing homogeneous, heterogeneous, and varied levels of response rate distribution across cohorts.
	
	In section 2.1 the methods used in this paper are presented in detail. In section 2.2 the setting, the simulation methodology and the explored scenarios are discussed, including the approach to compare the results. In section 3, the results from our simulation study are presented as the evaluation of the methods in terms of bias, MSE and amount of information borrowing. The methodology and the limitations of the comparative evaluation are discussed in the Discussion section of this paper, where we also provide suggestions for trialists. 

\section{Methods}
%%%%%%%%%%%%%%%%%%%%%%%%%%%%%%%%%%%%%%%%
\subsection{Estimators}
    The objective of estimation in the trial is to produce, for each cohort $i$, an estimate $\hat{p}_i$ of the \emph{true response rate} $p_i$ of that cohort, i.e. the probability of responding to the treatment for a randomly chosen patient in the $i^{th}$ cohort. Some methods assess $p_i$ through the the log odds parameter $\theta_i$ defined as $\theta_i = \log(\frac{p_i}{1-p_i})$, or, equivalently, $p_i = \frac{1}{1 + \exp(-\theta_i)}$.
    
    In this paper, besides the cohort-specific sample proportion (the observed number of responders divided by the total number of subjects), seven Bayesian procedures to estimate cohort-specific response rates allowing for borrowing are evaluated:

\subsubsection*{Estimator based on Berry et al.\cite{Berry} (2013)}
    Berry et al. \cite{Berry} (2013) discussed the use of a Bayesian hierarchical model (BHM) as a method that allows for borrowing information across all cohorts. The BHM assumes that the log odds parameters $\theta_i$ are exchangeable between cohorts,\footnote{The methods Berry et al. \cite{Berry}, EXNEX \cite{EXNEX}(exchangeable-non exchangeable) and Jin et al. \cite{Jin} model the response rate $p_i$ using the log odds transformation. Instead of modelling $p_i$ directly, they model the distribution of $\theta_i = logit(p_i) = log(\frac{p_i}{1 - p_i})$.} meaning that all cohorts follow the same distribution i.e. $N(\mu, \sigma^2)$. The hyper-priors, for $\mu$ and $\sigma^2$, are defined as a normal distribution $N(\mu_0, \sigma^2_0)$ and an inverse gamma distribution $\sigma^2 \sim IG(\lambda_1, \lambda_2)$ respectively. 

 \subsubsection*{Estimator based on Neuenschwander et al.\cite{EXNEX}(2016)}
    The EXNEX \cite{EXNEX}(exchangable-nonexchangable) method is a BHM method that extends the conventional Berry’s BHM by relaxing the assumption of all cohorts being exchangeable. Less information is being borrowed between non-similar cohorts. The log odds parameter $\theta_i$ for each cohort follows either a distribution which allows to exchange information, EX: $\theta_i  \sim N(\mu_0,\sigma^2_0)$ with probability $w$, or a distribution that is non-exchangeable with a probability 1- $w$, NEX: $\theta_i  \sim N(m_i,v^2_i)$. The hyper-parameters employed in the NEX component, namely $m_i, v_i^2$ and $w$, are fixed. In this study, priors and parameters were specified in accordance with the recommendations of the EXNEX \cite{EXNEX} authors. The hyper-parameters used in the EX component are $\mu_0 \sim N(0,10)$ and $\sigma^2_0 \sim \textrm{half-normal}(1)$. For the purposes of this paper the ‘bhmbasket’ R package is used for the calculation of the estimate.   

\subsubsection*{Estimator based on Psioda et al.\cite{Psioda} (2019)}
    Here, response rates are estimated following the procedure in Psioda et al. \cite{Psioda}, who propose a Bayesian model averaging approach. All possible models (ranging from the most parsimonious model in which all estimates are constrained to be equal to the most complex model in which all estimates are allowed to differ) are assigned a prior probability of being true. In addition, a beta prior is used for the response rate estimate in each cohort. Based on the observed data, the posterior model probabilities and model-specific posterior distributions for the response rates are determined. Cohort-specific response rates are calculated as the weighted average of the mean of the model-specific beta posterior distributions, with weights equal to the posterior model probabilities. Estimates were obtained using the function `bma` (version 0.1.2) in the R package `bmabasket`. We use the default parameter specifications, with the exception of the prior for the response rates, which are defined to be uniform (Beta(1,1)), instead of weakly informative. 

\subsubsection*{Estimator based on Fujikawa et al. \cite{Fuji}}
    The Fujikawa et al. \cite{Fuji} method is a Bayesian approach that borrows information across cohorts based on the similarity of their response rates. First, a uniform $(Beta(1,1))$ prior is used for the response rate in each cohort, which is updated based on the observed data. The pairwise similarity between the resulting posterior Beta distributions is then determined using the Jensen-Shannon divergence. If the similarity exceeds a given pre-specified threshold (denoted by $\tau$), the posteriors are ‘combined’ (i.e. borrowing will take place, by updating the parameters of the posterior distribution for a given cohort as a ‘weighted average’ of the initial posterior with all the other cohorts with similar effect size). We use the means of the possibly updated posterior distributions as the point estimates for the response rates. 

\subsubsection*{Estimator based on Jin et al. \cite{Jin} (2020)}
    The Bayesian hierarchical model with a correlated prior (CBHM) proposed by Jin et al. \cite{Jin} (2020) is a method that allows borrowing more information between possibly homogeneous cohorts and less when the treatment effect seems heterogeneous. For the log odds parameter $\theta_i$, it is assumed that it is specified as: $\theta_i = \theta_0 + \eta_i + \epsilon_i$. The $\eta_i$ are cohort-specific effects which follow a multivariate normal distribution with correlation matrix $\Omega$, $\eta_i \sim MVN(0, \sigma^2 \Omega)$. The similarity between two cohorts is identified in the $\Omega$ matrix, a correlation function is generated by the pairwise distance measures $d_{ij}$. Three different distance measures are considered in the original paper (the Kullback-Leibler distance (KL), the Hellinger (H) distance and the Bhattacharyya (B) distance). For the purposes of this paper, we will use the H distance. 

\subsubsection*{Estimator based on Chen \& Lee \cite{Lee} (2020)}
    Chen \& Lee\cite{Lee} proposed a Bayesian cluster hierarchical model. A two-step procedure, first the Chinese restaurant process (CRP) \cite{CRP} identifies the partitioning into clusters of each cohort using a non-parametric Dirichlet process mixture model (DPM). The values of the clustering matrix $C_{ij}$ are the proportion of two cohorts being classified in the same cluster. The second step is to use a Bayesian hierarchical model to estimate the log odds of the response rate $\theta_i \sim N( \mu_1, \frac{1}{\tau_1 m_i}$), given the cluster structure. The $m_i$ in this model is set to $C_{ij}$ for the specific subgroup $i$. A hyper-prior for the mean $\mu_1$ follows a normal distribution $N(\mu_2, \frac{1}{\tau_2})$, a hyper-prior for the precision controlling the amount of borrowing between cohorts $i$ and $j$, $\tau_1 \sim Gamma(\alpha_1,\beta_1)$ and $\mu_i$ is the respective model indicated by the $C_{ij}$ matrix. The variance of the hyper-prior was explored by Chen \& Lee \cite{chen_lee_old}, $\tau_2 = 0.1$. The clustering matrix probability of subgroup $i$ and $j$ to share information influences the variance of the $\theta_i$ distribution. As the similarity value increases, the information sharing becomes larger and the posterior distribution variance decreases. The ‘BCHM’ R package is used for the response rate calculation given the default parameters proposed by the authors. (More details in Appendix B)

\subsubsection*{Estimator based on Liu et al.\cite{Liu_kane} (2022)}
    Liu et al. \cite{Liu_kane} propose a method that evaluates the probabilities of all possible models (referred to as 'partitions'), assigning a prior probability to each partition differently than Psioda et al. \cite{Psioda}. The choice of prior is carefully discussed, with a parameter, delta, introduced to determine the level of influence each model exerts. The authors suggest values of 0 (uniform prior), 1, and 2 for delta. Unlike the Psioda method, which uses a weighted average, Liu et al.\cite{Liu_kane} select the most probable partition to compute the pairwise similarity matrix among cohorts. This matrix is then used to calculate the parameters of the Beta posterior distribution. In the local multiple exchangeability model (local MEM), this similarity matrix is used to determine which baskets are grouped together based on the highest posterior probability partition. Information borrowing is then carried out locally within these identified groups. Specifically, baskets that belong to the same block, as defined by the selected partition, are treated as exchangeable, and data from these baskets are used to update the parameters of their Beta posterior distributions. This localized borrowing approach allows for precise information sharing while respecting differences between dissimilar baskets.

\subsection{Simulation study}

    \subsubsection*{Setup}
    In the simulation study, we construct a basket trial with six different cohorts, which we consider to be a realistic choice. For simplicity, all cohorts of the study are assumed to be single-stage and have equal size. Limited to no prior knowledge is assumed to be available. The total number of patients per cohort is denoted by $n_i$ and the number of responses is indicated by $r_i$. 

\subsubsection*{Scenarios}

    The scenarios in which we evaluate the response rate estimators differ in the distribution of the true response rates $p_i$ across cohorts and the number of patients per cohort. We classify these scenarios into two types: homogeneous and heterogeneous (table \ref{tab:confint}). The choice of the different scenarios is based on the methodological aspects of the methods, but also extreme practical examples, like the KEYTRUDA \cite{KEYTRUDA} trial, lead us to consider a broad range of scenarios. \\
    
\begin{itemize}
    \item \textbf{Homogeneous Scenarios:}
    
    In homogeneous scenarios, all cohorts have similar or identical true response rates. These scenarios are designed to evaluate how well the estimators perform when there is little to no variation between cohorts. Specifically:
    
    \textbf{Scenarios 1.A.1 to 1.A.3}: The true response rate is the same across all six cohorts, with values set at 0.1, 0.3, or 0.5, respectively.\\

     \item \textbf{Heterogeneous Scenarios:}
    
    In heterogeneous scenarios, the cohorts have more distinct response rates, representing a wider range of treatment effects across the different groups. These scenarios are designed to test the ability of the estimators to handle substantial variability:\\

    \textbf{Scenarios 1.B.1 to 1.B.4}: Small variations in true response rates are introduced between cohorts. In scenarios 1.B.1 and 1.B.2, the response rates differ slightly (by 0.025 and 0.05 respectively) around an overall mean of 0.5. In scenarios 1.B.3 and 1.B.4, the overall mean is 0.3, with variations of 0.025 and 0.05 respectively.\\
        
    \textbf{Scenarios 2.A.1 to 2.D.3}: These scenarios represent situations where two distinct groups of cohorts have different response rates. Scenarios 2.A.1 to 2.B.3 response rate with difference of 0.2. Scenarios 2.C.1 to 2.C.3 represent a larger response rate difference of 0.4, while scenarios 2.D.1 to 2.D.3 consider an even greater difference of 0.6 between the two groups.\\
    
    \textbf{Scenarios 3.A.1, 3.A.3, 3.B.1, and 3.B.3}: In  scenarios, 3.A.1, 3.A.3, 3.B.1, 3.B.3, we assume three different response rates, where four of the cohorts have the same response rate, while the remaining two cohorts have different rates. Scenarios 3.A.2 and 3.B.2 describe cases where three pairs of cohorts each have the same response rate, resulting in three different groups.\\
    
    The responses of subjects in each cohort were generated from a binomial distribution with true probability $p_i$, where $i \in \lbrace 1,2,...,6 \rbrace$. The number of subjects per cohort was assumed to be $N = \lbrace 10, 20, 30, 100 \rbrace$.
    
    Details concerning the specification of prior distribution parameters and tuning parameters of the seven estimators are provided in Appendix B. R-code is available as online supplementary material.
    
\end{itemize}

\begin{table}[h]
\begin{center}
\begin{tabular}{ccccccccccccccc}
\hline
 & Scenarios && Coh A && Coh B  && Coh C  && Coh D  &&  Coh E && Coh F\\ 
  \hline
 &  1.A.1 &&   0.1    && 0.1    && 0.1    && 0.1    && 0.1    && 0.1\\ 
 &  1.A.2 &&   0.3    && 0.3    && 0.3    && 0.3    && 0.3    && 0.3\\ 
 &  1.A.3 &&   0.5    && 0.5    && 0.5    && 0.5    && 0.5    && 0.5\\ 
   \hline
 &  1.B.1 &&   0.4375 && 0.4625 && 0.4875 && 0.5125 && 0.5375 && 0.5625\\ 
 &  1.B.2 &&   0.375  && 0.425  && 0.475  && 0.525  && 0.575  && 0.625\\ 
 &  1.B.3 &&   0.2375 && 0.2625 && 0.2875 && 0.3125 && 0.3375 && 0.3625\\ 
 &  1.B.4 &&   0.175  && 0.225  && 0.275  && 0.325  && 0.375  && 0.425\\ 
 &  2.A.1 &&   0.3    && 0.5    && 0.5    && 0.5    && 0.5    && 0.5\\ 
 &  2.A.2 &&   0.3    && 0.3    && 0.3    && 0.5    && 0.5    && 0.5\\ 
 &  2.A.3 &&   0.3    && 0.3    && 0.3    && 0.3    && 0.3    && 0.5\\ 
 
 &  2.B.1 &&   0.1    && 0.3    && 0.3    && 0.3    && 0.3    && 0.3\\ 
 &  2.B.2 &&   0.1    && 0.1    && 0.1    && 0.3    && 0.3    && 0.3\\ 
 &  2.B.3 &&   0.1    && 0.1    && 0.1    && 0.1    && 0.1    && 0.3\\ 
 
 &  2.C.1 &&   0.1    && 0.5    && 0.5    && 0.5    && 0.5    && 0.5\\ 
 &  2.C.2 &&   0.1    && 0.1    && 0.1    && 0.5    && 0.5    && 0.5\\ 
 &  2.C.3 &&   0.1    && 0.1    && 0.1    && 0.1    && 0.1    && 0.5\\ 
  
 &  2.D.1 &&   0.1    && 0.7    && 0.7    && 0.7    && 0.7    && 0.7\\ 
 &  2.D.2 &&   0.1    && 0.1    && 0.1    && 0.7    && 0.7    && 0.7\\ 
 &  2.D.3 &&   0.1    && 0.1    && 0.1    && 0.1    && 0.1    && 0.7\\ 

 &  3.A.1 &&   0.1    && 0.4    && 0.7    && 0.7    && 0.7    && 0.7\\ 
 &  3.A.2 &&   0.1    && 0.1    && 0.4    && 0.4    && 0.7    && 0.7\\ 
 &  3.A.3 &&   0.1    && 0.1    && 0.1    && 0.1    && 0.4    && 0.7\\ 
   
 &  3.B.1 &&   0.1    && 0.4    && 0.9    && 0.9    && 0.9    && 0.9\\ 
 &  3.B.2 &&   0.1    && 0.1    && 0.4    && 0.4    && 0.9    && 0.9\\ 
 &  3.B.3 &&   0.1    && 0.1    && 0.1    && 0.1    && 0.4    && 0.9\\ 
 
\hline
\end{tabular}
\end{center}
\caption{Scenarios used in the explored simulations. The first 3 scenarios are the Homogeneous scenarios and the rest are referred to the Heterogeneous scenarios.}
\label{tab:confint}
\end{table}

    \subsubsection*{Evaluation criteria}

    The estimators will be compared based on the average absolute bias and average MSE, as defined in Table \ref{tab:measure}.

\begin{table}[h]
\begin{center}
\begin{tabular}{ccccc}
\hline
 &  && Formula \\ 
\hline
 & MeanAbsBias    & $=$&	$\frac{\sum^6_{i=1} |E(\hat{p}_i) - p_i |}{6} $\\
 &&&\\
 & $Mean MSE$		& $=$&  $\frac{\sum^6_{i=1}[Bias(\hat{p}_i)^2 + Var(\hat{p}_i)]}{6} $\\
  &&&\\
& $Shrinkage$	& $=$&  $1 - \frac{max(E(\hat{p_1}), ... , E(\hat{p_6})) - min(E(\hat{p_1}), ... , E(\hat{p_6})) }{max(p_1, ... , p_6) - min(p_1, ... , p_6)}$ \\
\hline
\end{tabular}
\end{center}
\caption{Evaluation criteria}
\label{tab:measure}
\end{table}

An additional measure used to provide further insight in the results, is the Shrinkage to the total mean, which is defined as the difference between the maximum and the minimum estimated response rate, divided by the simulated max and min. We use the form as presented in table \ref{tab:measure}, such that the closer this value is to 1, the greater the shrinkage. When the shrinkage is close to 0, it indicates that the methods are not borrowing much information.

\section{Results} 
  
%%%%%%%%%%%%%%%%%%%%%%%%%%%%%%%%%%%%%%%%

\subsection{Homogeneous scenario}
\begin{figure}[h]
\centering
\includegraphics[width=17cm,height=14cm]{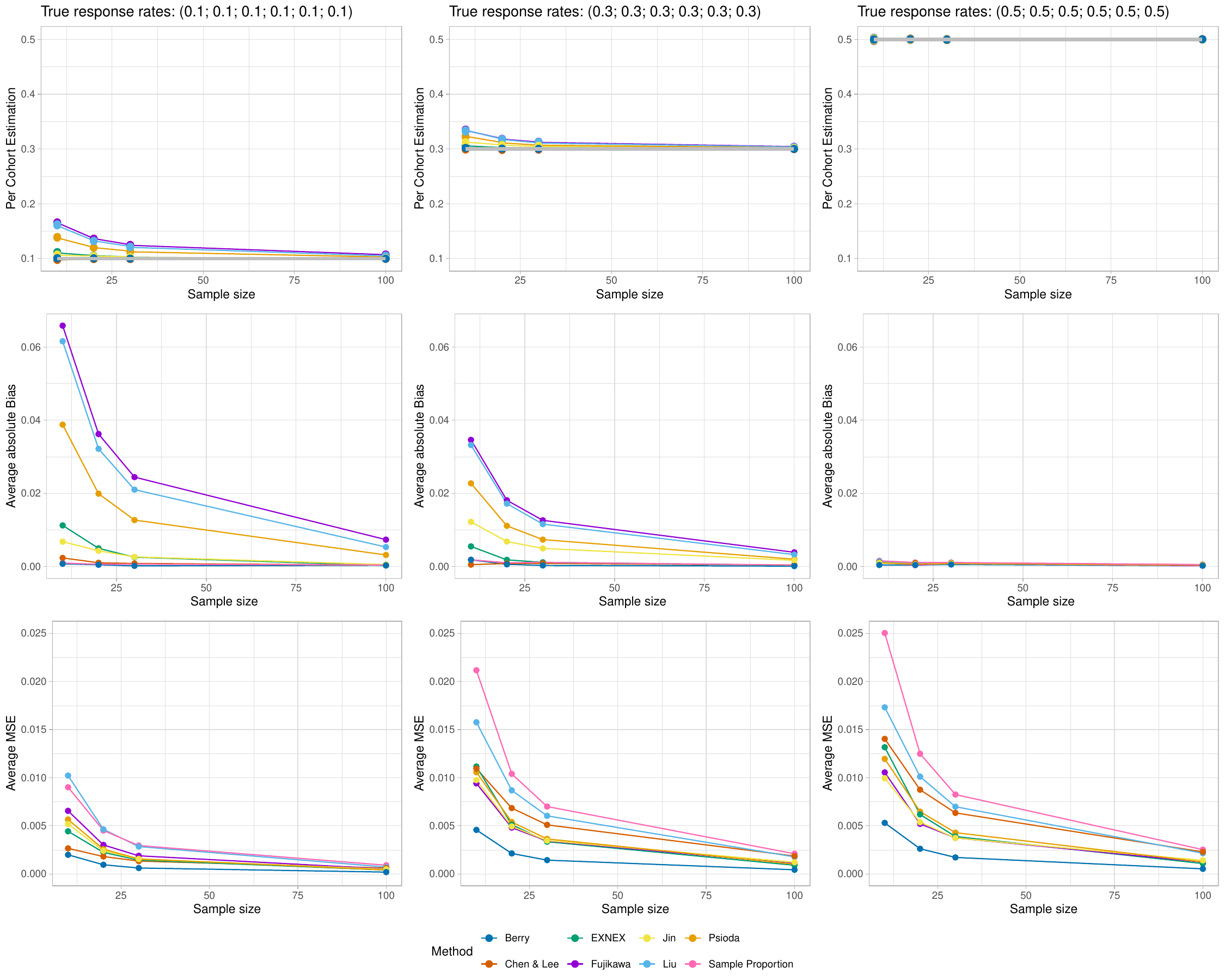}
\caption{\textit{Estimates on scenarios of the same true response rate 1.A.1 to 1.A.3 in 4 sample size points, 10, 20, 30, 100 (patients per cohort). In the 1st row the estimates are presented, in the second and third, the average absolute bias and the average MSE of the respective scenarios. The gray lines in the first row reflect the true response rates. The prior distribution in this graph is having a mean of the 0.5}}
\label{fig:same_6coh}	
\end{figure}

    In the homogeneous scenarios with exactly equal true response rates across the baskets (scenarios 1.A.1 to 1.A.3, fig \ref{fig:same_6coh} all Bayesian estimators except Chen \& Lee show on average positive bias for response rates 0.1 and 0.3, but not when the response rate is 0.5. The Chen \& Lee shows negative bias when the true estimates are lower than 0.5. Among the Bayesian estimators, Berry's BHM \cite{Berry} shows the smallest bias and average MSE, regardless of the sample size per basket. The EXNEX and Jin methods are slightly shifting towards 0.5. When the sample size is small, the average absolute bias for some of the other estimators, particularly those based on Liu and Fujikawa, appears quite substantial towards 0.5 when the true common response rate is 0.1, but overall bias decreases to negligible when the true response rate gets closer to 0.5. \\

\begin{figure}[h]
\centering
\includegraphics[width=17cm,height=15cm]{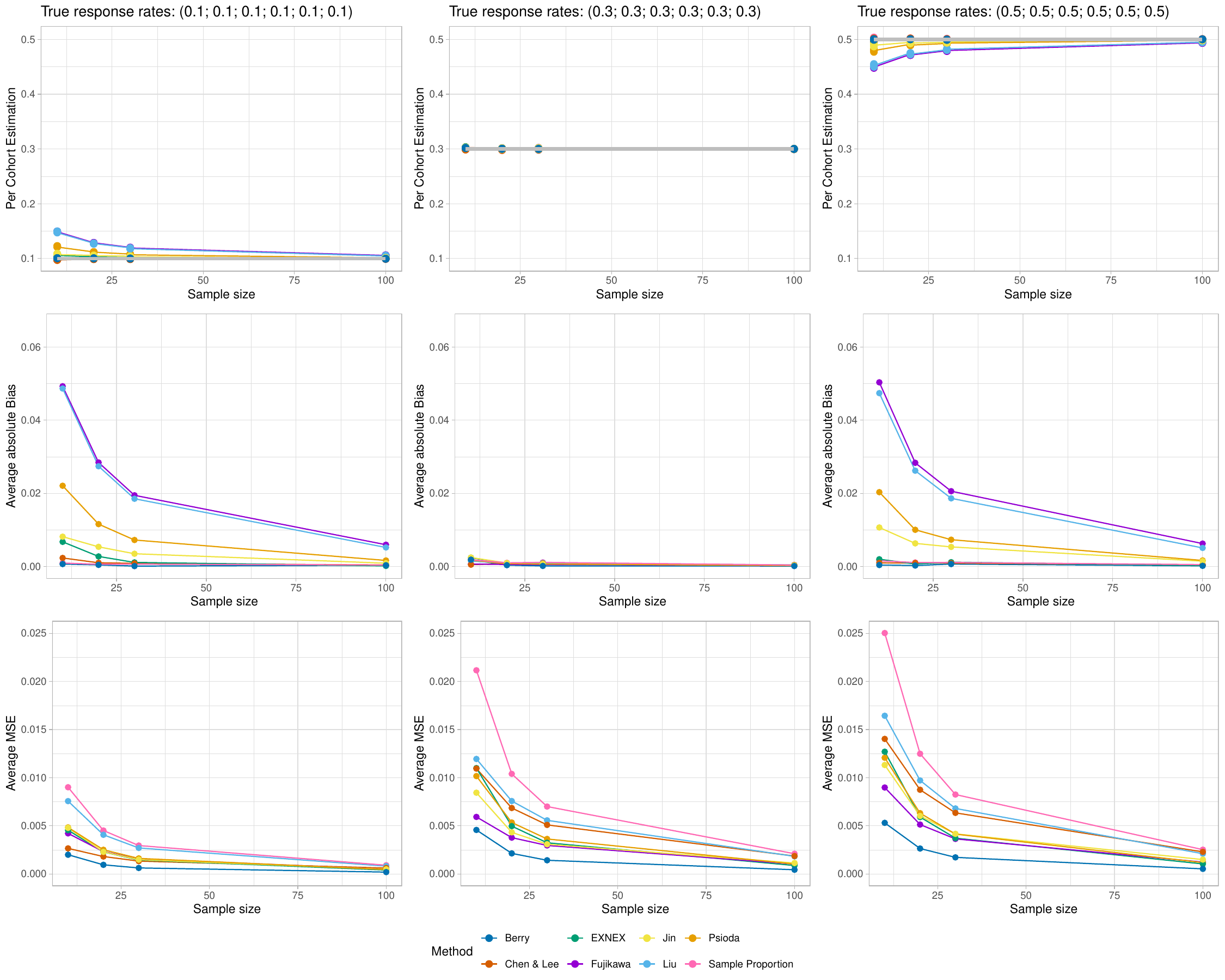}
\caption{\textit{Estimates on scenarios of the same true response rate 1.A.1 to 1.A.3 in 4 sample size points, 10, 20, 30, 100 (patients per cohort). In the 1st row the estimates are presented, in the second and third, the average absolute bias and the average MSE of the respective scenarios. The gray lines in the first row reflect the true response rates. The prior distribution in this graph is having a mean of the 0.3}}
\label{fig:same_03_prior}	
\end{figure}

    To determine whether this effect is actually due to the prior choices, we conducted additional simulations, (Appendix C, fig \ref{fig:same_03_prior},\ref{fig:01_03_03prior}) in which the parameters of each method were tuned to have a prior mean of 0.3. The results indicate that the prior mean plays a role and influences whether the estimates overestimate or underestimate the effect, towards to the chosen prior mean. Berry's method is not affected by the prior, as well as Chen \& Lee. The EXNEX and Jin methods are affected slightly. The Psioda method is affected, but less than the Fujikawa and Liu methods. 

\clearpage

\subsection{Heterogeneous Scenario}
\begin{figure}[h]
\centering
\includegraphics[width=17cm,height=15cm]{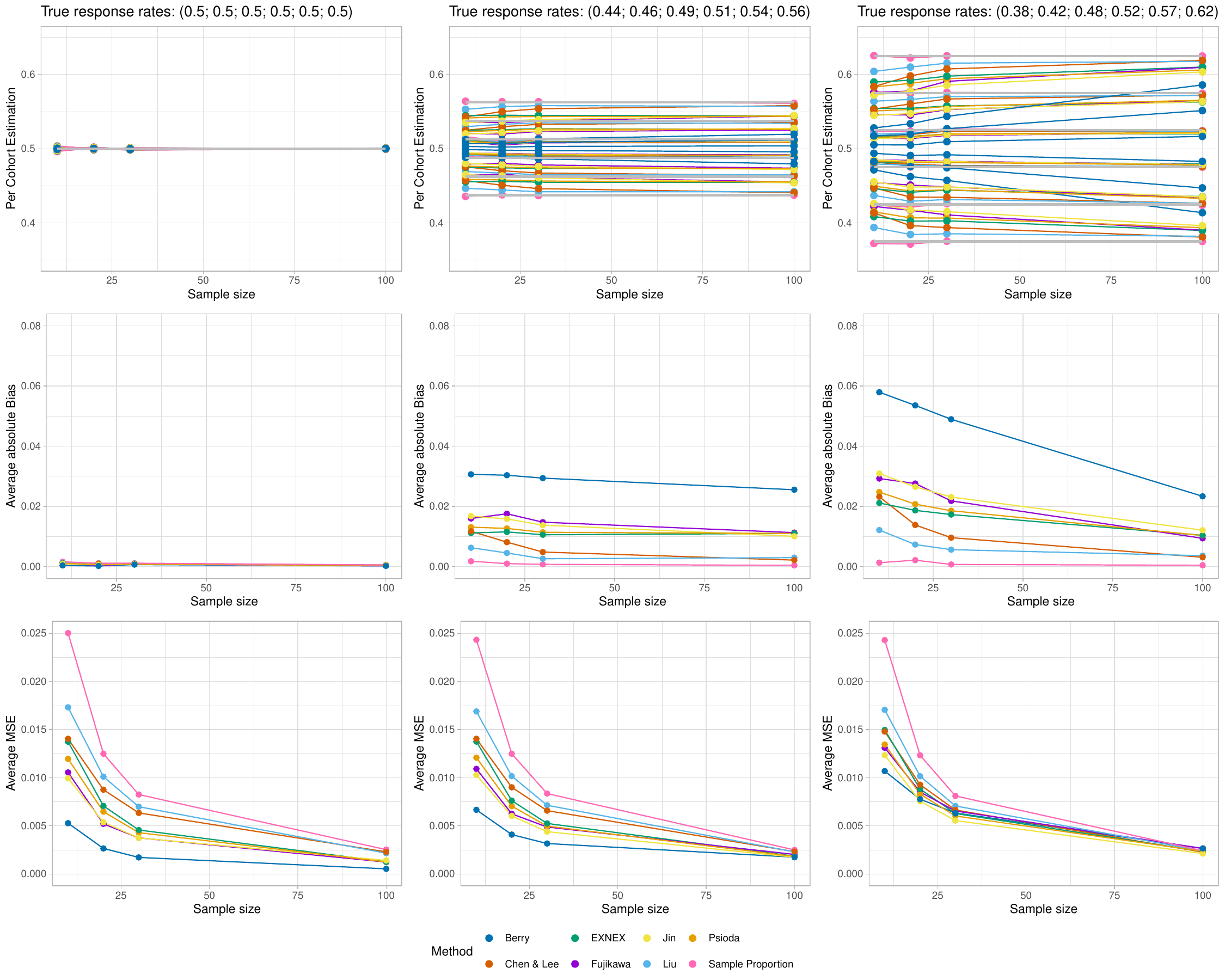}
\caption{\textit{Estimates on scenarios 1.A.3 to 1.B.2, in 4 sample size points, 10, 20, 30, 100, in the 1st row. In the second row, is the average absolute Bias of the respective scenarios and the third row is the average of MSE.} }
\label{fig:same_var_6coh}	
\end{figure}

\begin{figure}[h]
\centering
\includegraphics[width=17cm,height=15cm]{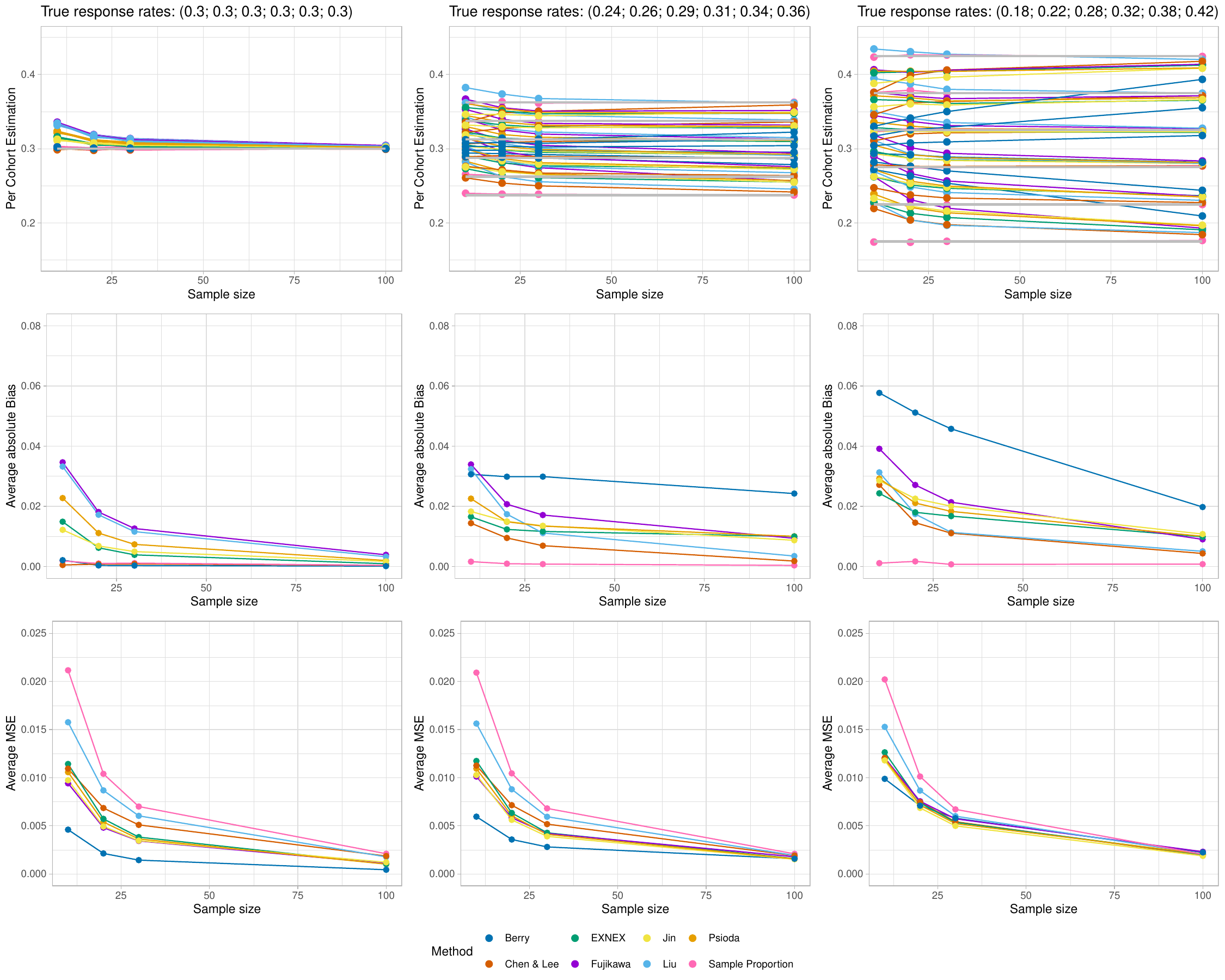}
\caption{\textit{Estimates on scenarios 1.A.2, 1.B.3 and 1.B.4, in 4 sample size points, 10, 20, 30, 100, in the 1st row. In the second row, is the average absolute Bias of the respective scenarios and the third row is the average of MSE.} }
\label{fig:same_var_03_6coh}	
\end{figure}

    If we allow for some heterogeneity in the true RRs of the six cohorts (scenarios 1.B.1 to 1.B.4, fig \ref{fig:same_var_6coh} and \ref{fig:same_var_03_6coh}), the observed bias of the different Bayesian estimators increases. The average mean squared errors of the different estimators overall increase and become more similar.

\begin{figure}[h]
\centering
\includegraphics[width=17cm,height=15cm]{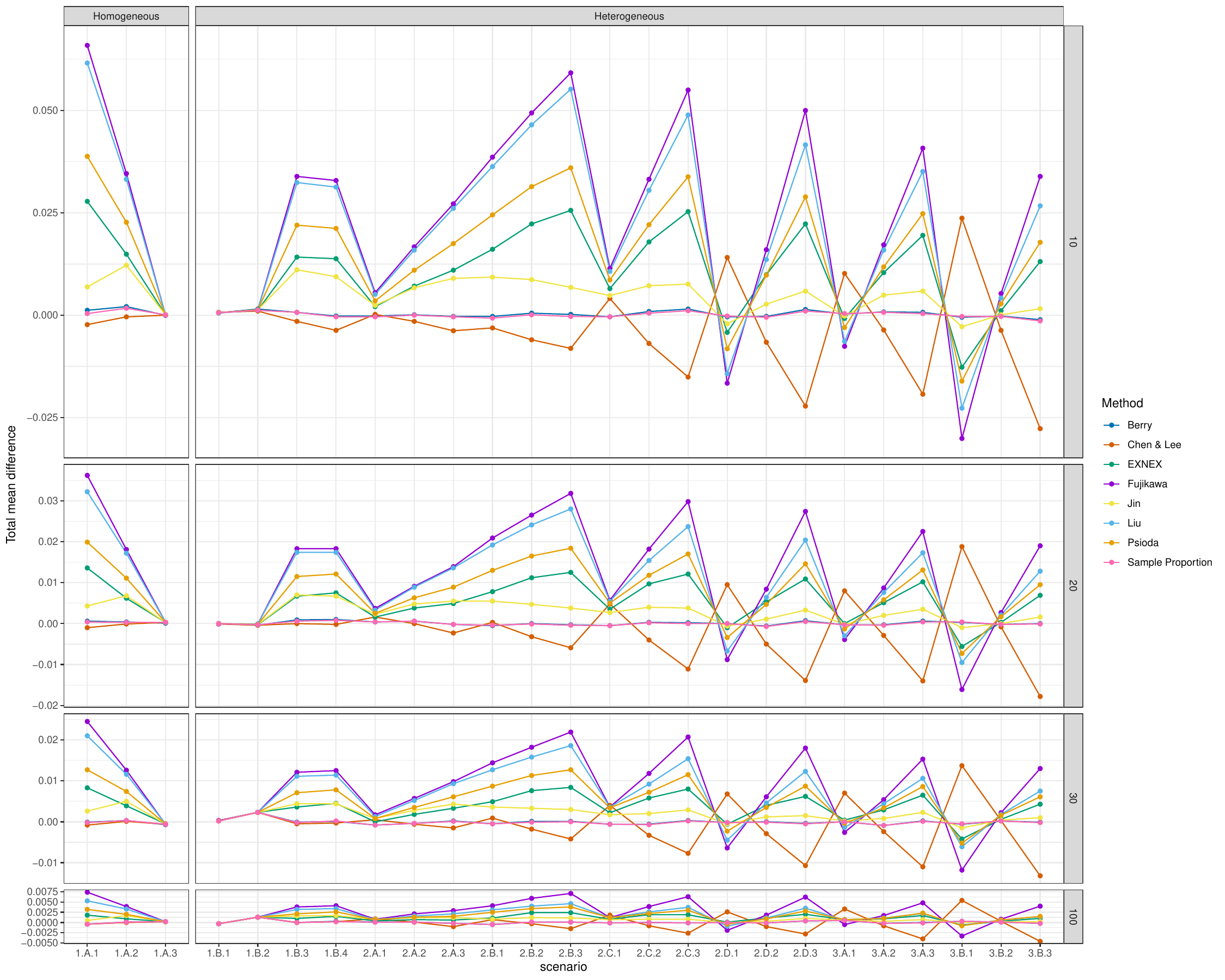}
\caption{\textit{Difference between the mean true RR of all cohorts with the mean of the simulated response rates in each method and every scenario. Each row indicates the different sample size points.} }
\label{fig:Total_est_mean}	
\end{figure}

    In fig \ref{fig:same_var_6coh}, in the first row the estimates can again be seen to be biased towards 0.5, regardless of the variation level. In fig \ref{fig:Total_est_mean} it can be seen that different true RRs across the 6 cohorts centered around 0.5, that the estimates shift to this mean of 0.5 of the cluster. In fig \ref{fig:same_var_03_6coh}, the mean of true RRs of the cohorts is 0.3, and most of the estimates show shift towards 0.5, which was not observed in scenarios where the overall cluster mean of the true RRs is 0.5.

    In these scenarios, Berry's BHM estimator has the lowest MSE when the response rate is less heterogeneous, compared to the other methods. However, with increasing variability, the average absolute bias appears relatively large, even with a large sample size. Also for other estimators, bias still appears to be present even with samples sizes as large as 100 per cohort.

\begin{figure}[h]
\centering
\includegraphics[width=17cm,height=15cm]{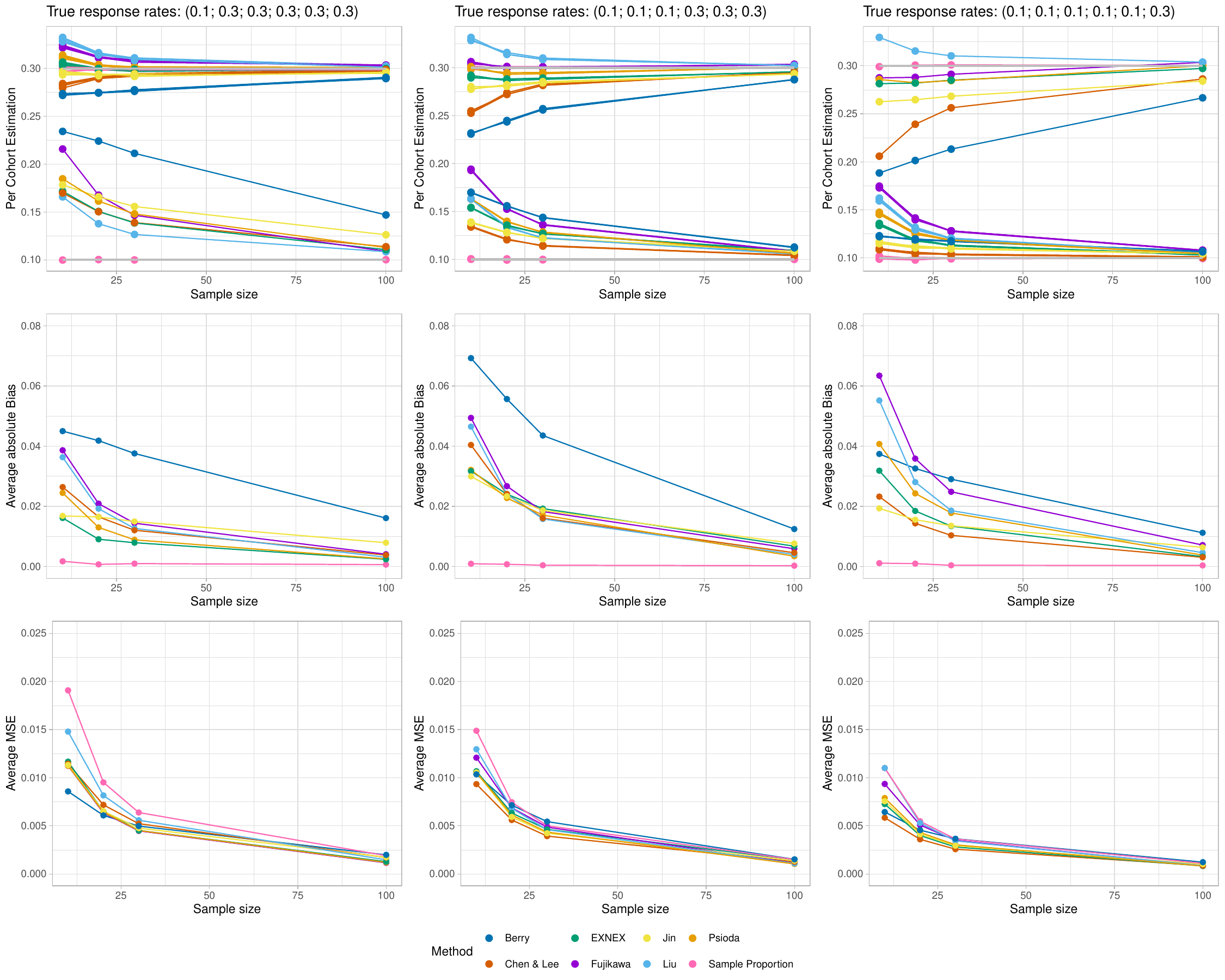}
\caption{\textit{Estimates on scenarios 2.B.1 to 2.B.3, in 4 sample size points, 10, 20, 30, 100, in the 1st row. In the second row, is the average absolute Bias of the respective scenarios and the third row is the average of MSE. The gray lines in the first row reflect the true response rates. The prior distribution in this graph is having a mean of the 0.5} }
\label{fig:01_03}	
\end{figure}

    When considering scenarios involving the grouping of cohorts (i.e., two or three distinct groupings), all methods effectively capture the structure (fig \ref{fig:01_03}). As anticipated, the sample proportion has the smallest average absolute bias (theoretically 0), but its average Mean Squared Error (MSE) is among the highest (see fig \ref{fig:01_03}). In this setting, Berry's Bayesian Hierarchical Model (BHM) estimator demonstrates the largest average absolute bias across most scenarios. Estimators following Fujikawa and Liu's approaches tend to shrink more towards the prior of 0.5, compared to other estimates. Jin's estimator shrinkage is small. EXNEX, Psioda, and Chen $\&$ Lee estimates also show a tendency towards the mean of 0.5, but with a notable exception for the Chen $\&$ Lee estimator (see fig \ref{fig:Total_est_mean}): it uniquely shifts towards 0 when the estimated total mean is below 0.5 and towards 1 when above 0.5.

\begin{figure}[h]
\centering
\includegraphics[width=17cm,height=15cm]{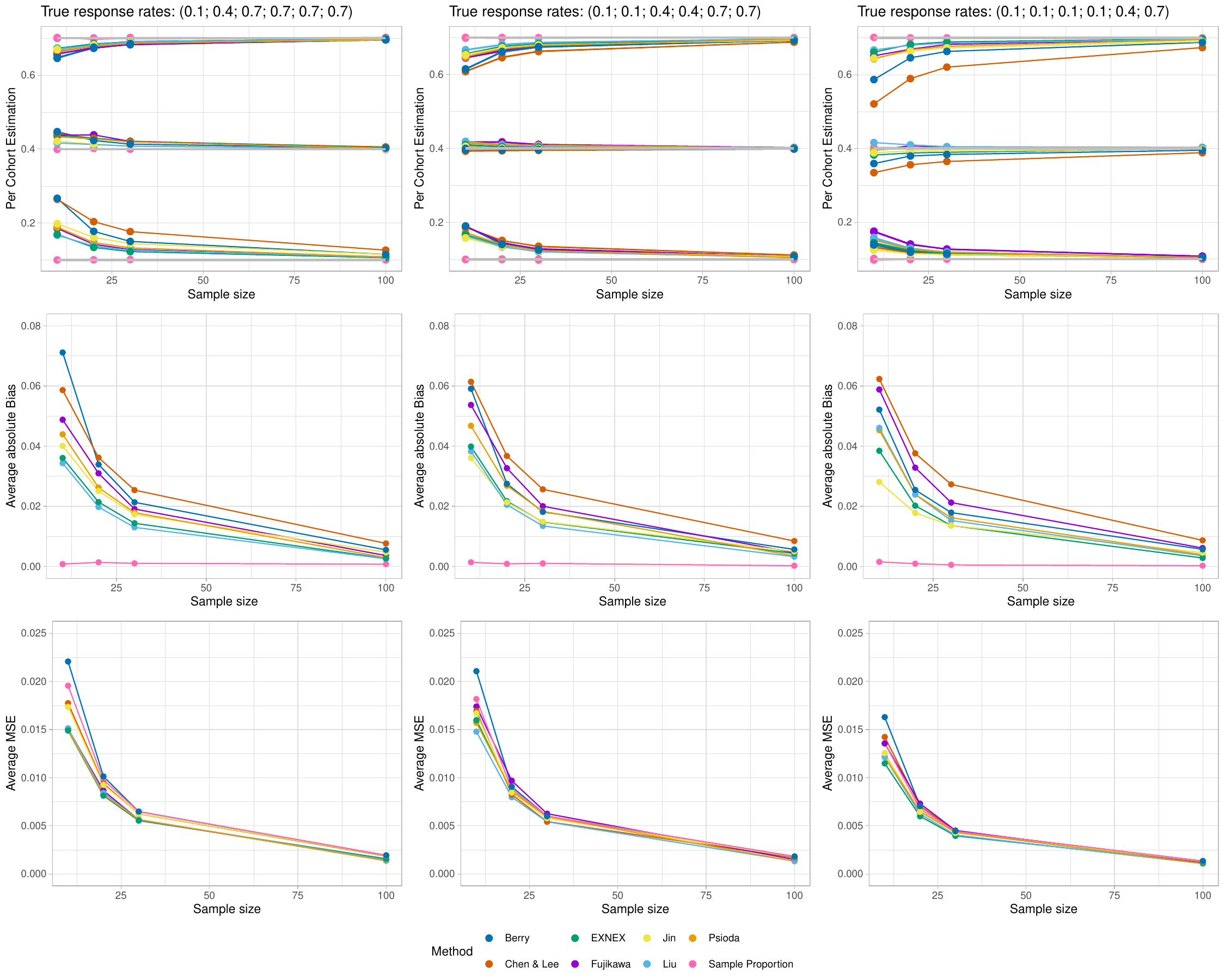}
\caption{\textit{Estimates on scenarios 3.A.1 to 3.A.3, in 4 sample size points, 10, 20, 30, 100, in the 1st row. In the second row, is the average absolute Bias of the respective scenarios and the third row is the average of MSE. The gray lines in the first row reflect the true response rates.} }
\label{fig:01_04_07}	
\end{figure}

\begin{figure}[h]
\centering
\includegraphics[width=17cm,height=15cm]{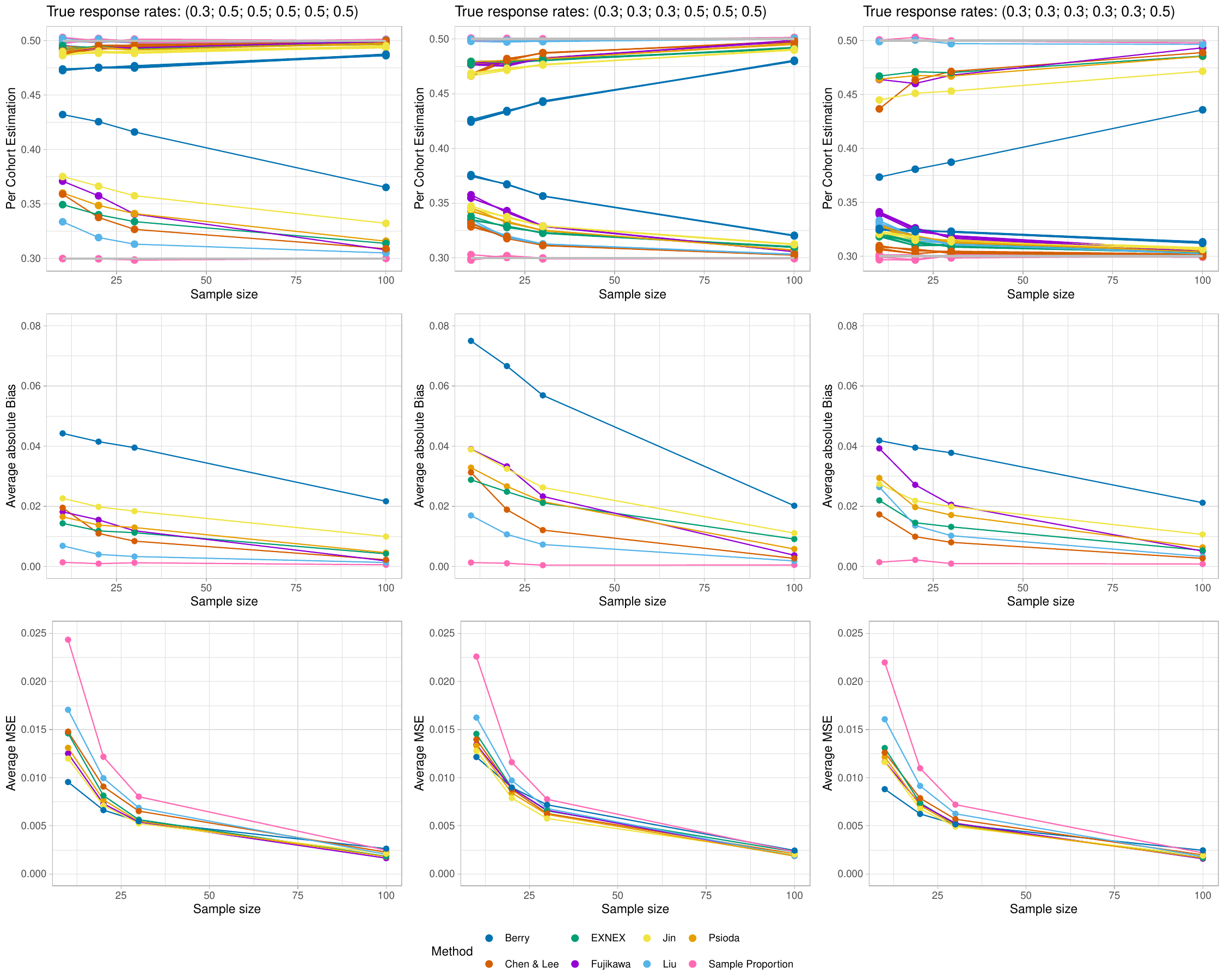}
\caption{\textit{Estimates on scenarios 2.A.1 to 2.A.3, in 4 sample size points, 10, 20, 30, 100, in the 1st row. In the second row, is the average absolute Bias of the respective scenarios and the third row is the average of MSE.} }
\label{fig:03_05_6coh}	
\end{figure}

\begin{figure}[h]
\centering
\includegraphics[width=17cm,height=15cm]{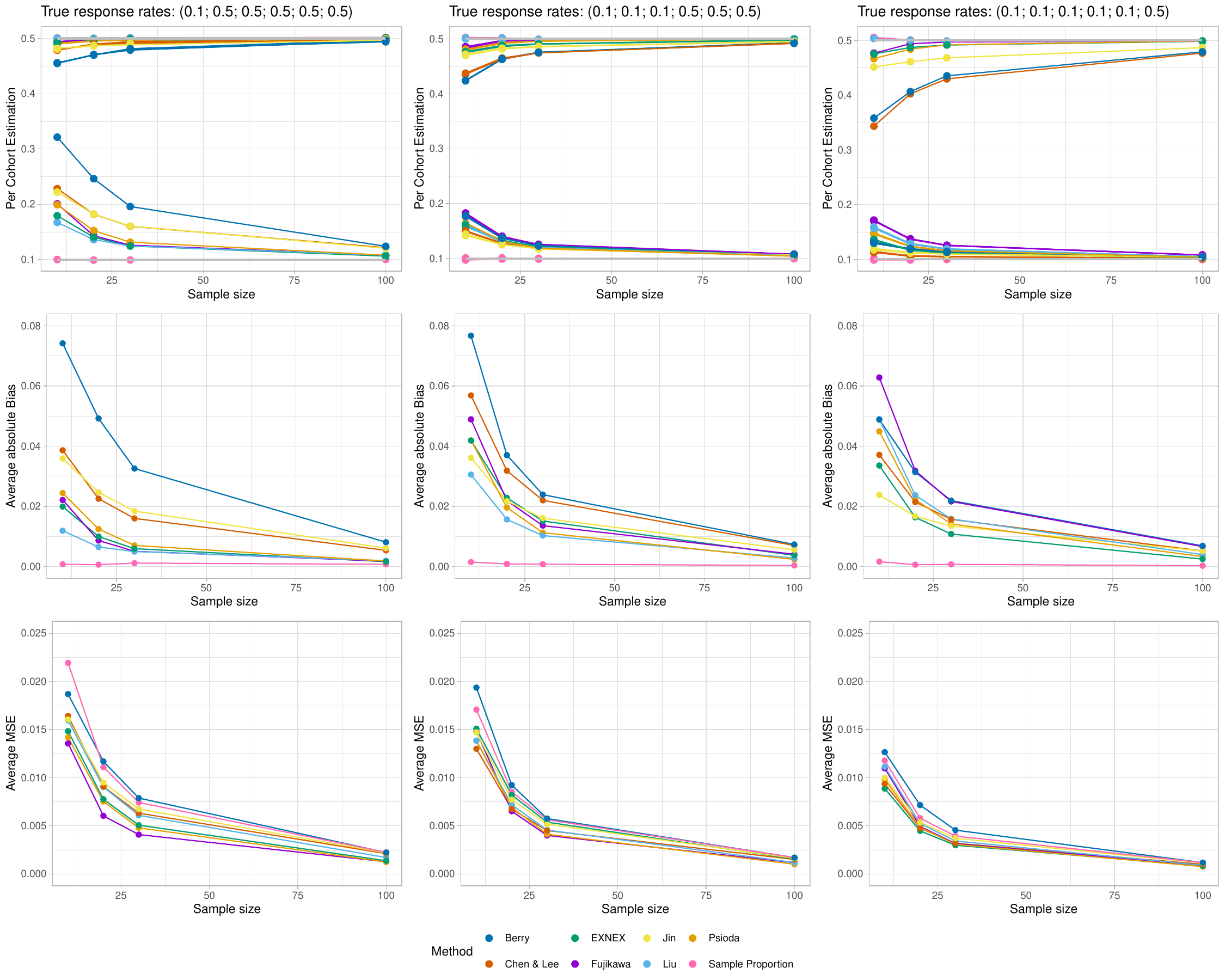}
\caption{\textit{Estimates on scenarios 2.C.1 to 2.C.3, in 4 sample size points, 10, 20, 30, 100, in the 1st row. In the second row, is the average absolute Bias of the respective scenarios and the third row is the average of MSE.} }
\label{fig:01_05_6coh}	
\end{figure}

\begin{figure}[h]
\centering
\includegraphics[width=18cm,height=15cm]{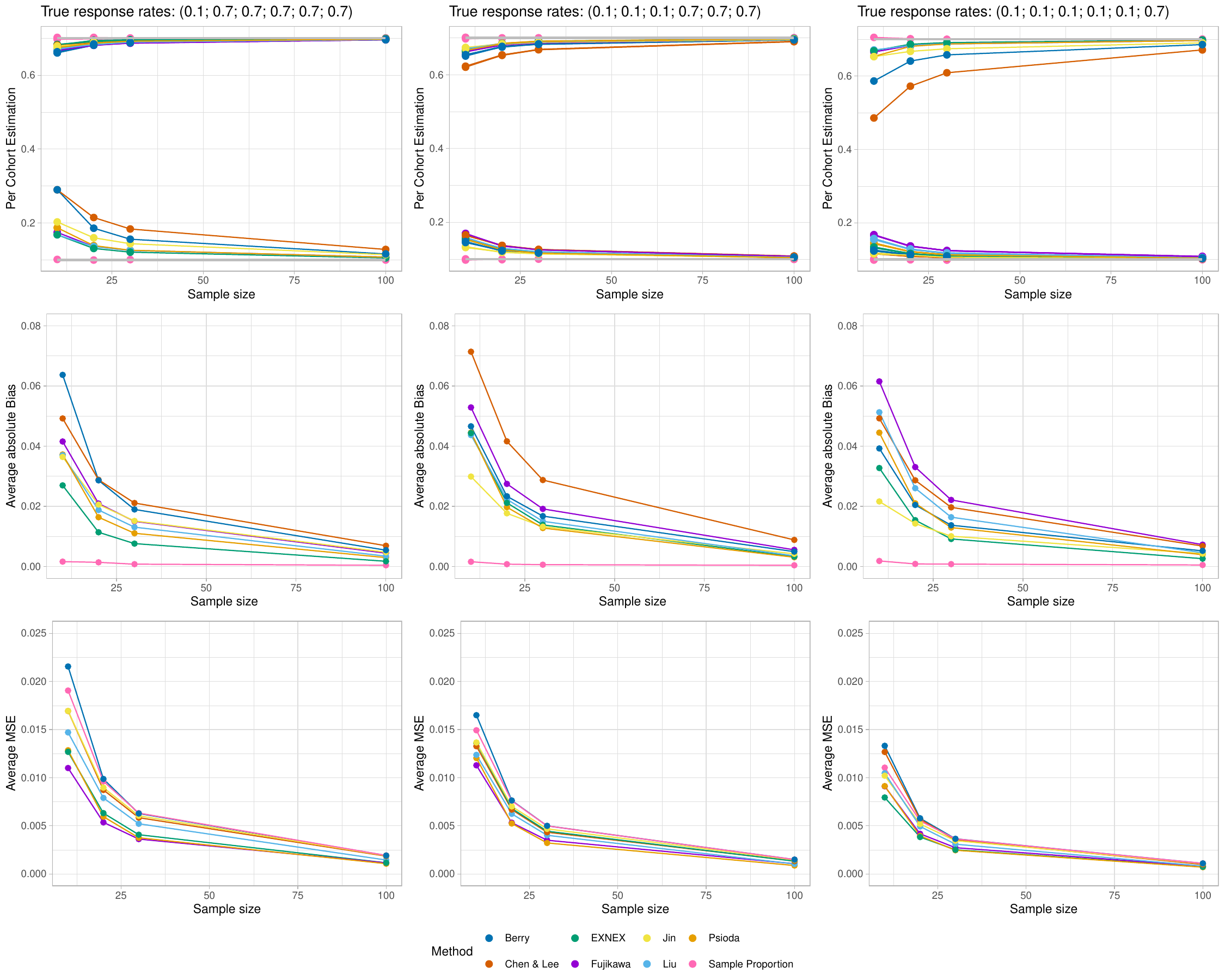}
\caption{\textit{Estimates on scenarios 2.D.1 to 2.D.3, in 4 sample size points, 10, 20, 30, 100, in the 1st row. In the second row, is the average absolute Bias of the respective scenarios and the third row is the average of MSE.} }
\label{fig:01_07_6coh}	
\end{figure}

    In scenarios as in fig \ref{fig:01_05_6coh}, where the actual response rate of five out of 6 cohorts is equal to 0.5 and one cohort deviates, the methods that tend to shift estimates towards 0.5 (such as Fujikawa’s) perform better in terms of MSE. Berry's BHM estimator appears favourable with respect to average MSE when the heterogeneity is small (see fig \ref{fig:03_05_6coh}, \ref{fig:same_var_03_6coh}). However, as the heterogeneity increases (see fig \ref{fig:01_05_6coh}, \ref{fig:01_07_6coh}, \ref{fig:01_04_07}), Berry's estimator's MSE grows relative to others. The observed patterns remain similar across all scenarios.

\begin{figure}[h]
\centering
\includegraphics[width=17cm,height=15cm]{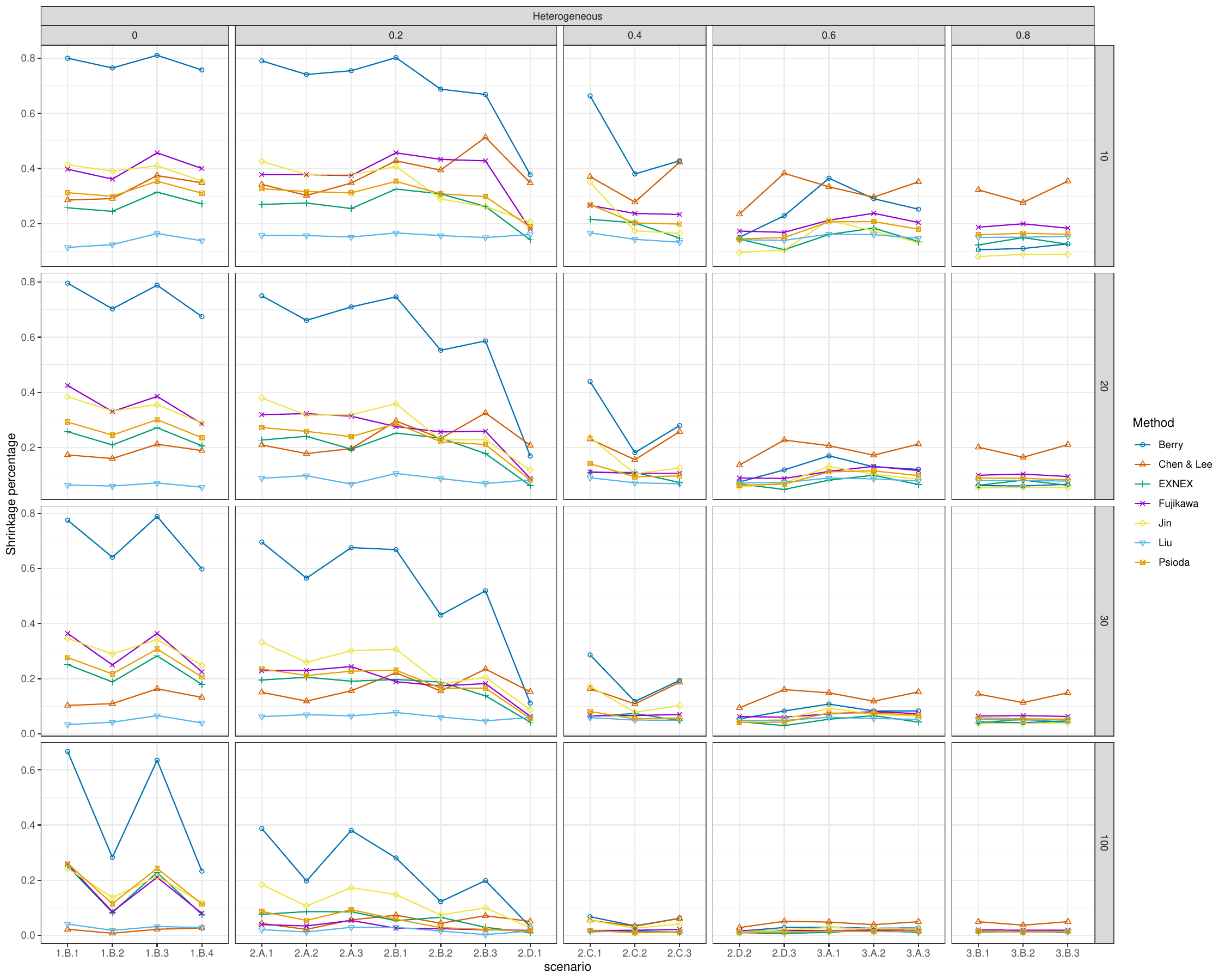}
\caption{\textit{Shrinkage of the estimates. Compare the range of the maximum and the minimum distance of the estimates with the true distance. Closer to 1 on y axis means the estimator shrinkage to the total mean is extreme. On the contrary if it's closer to 0, then the borrowing information is limited.} }
\label{fig:shrinkage}	
\end{figure}

    The diversity across different cohorts impacts the degree of shrinkage to the overall mean observed in our analysis. Specifically, as the heterogeneity between cohorts increases, we notice a corresponding decrease in shrinkage (see fig \ref{fig:shrinkage}). Berry's method consistently shrinks more towards the overall mean, irrespective of the level of heterogeneity. In contrast, Liu's method exhibits the least shrinkage across all scenarios. The Chen and Lee method tends to shrink towards the overall mean, similar to most methods when variability is low, but more prominently than others in scenarios with large between cohort heterogeneity. Overall, shrinkage decreases with greater cohort heterogeneity. Additionally, we observe that shrinkage tends to diminish as the sample size increases, although certain methods—like Berry's—maintain high shrinkage under conditions of small heterogeneity.

\begin{figure}[h]
\centering
\includegraphics[width=17cm,height=15cm]{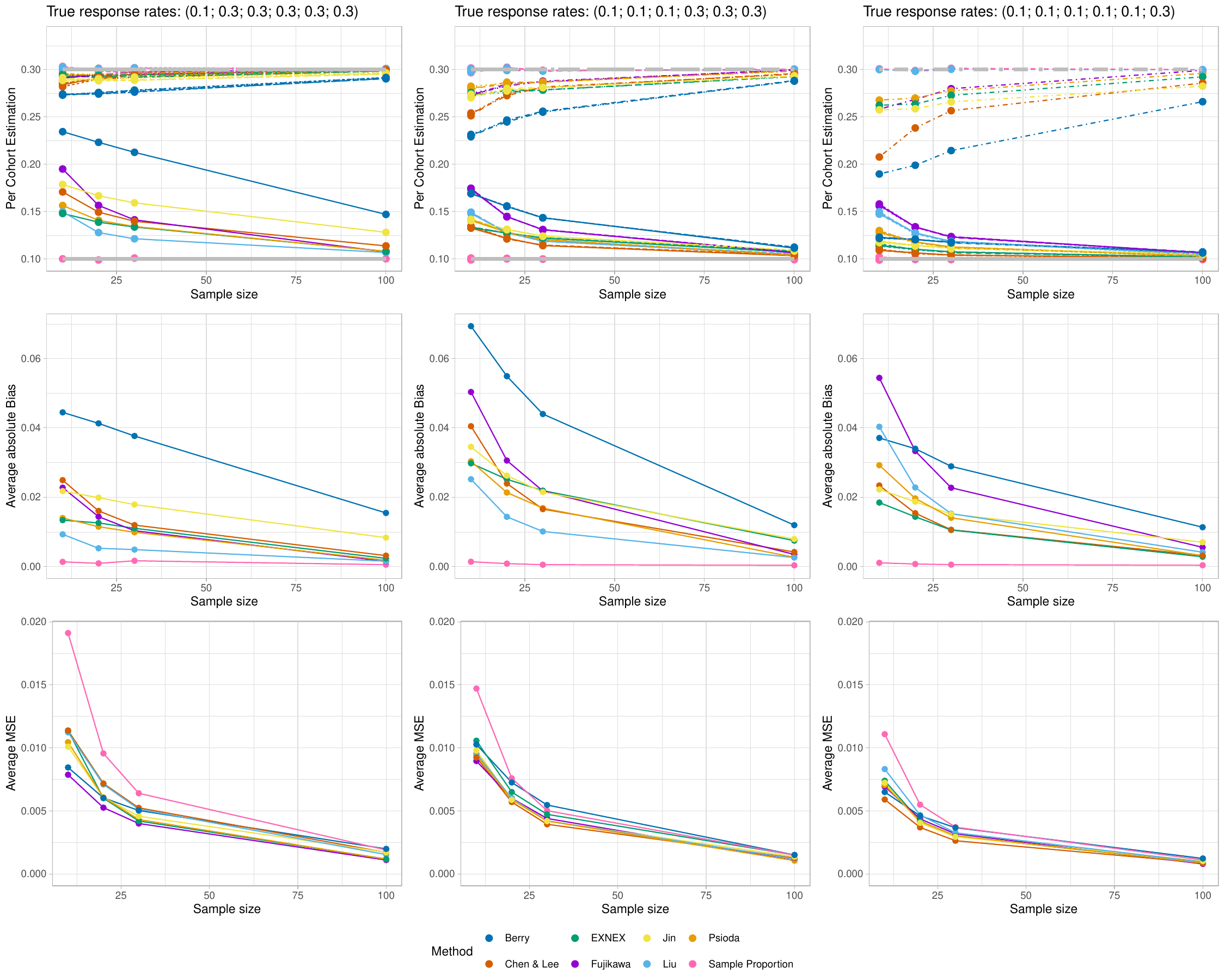}
\caption{\textit{Estimates on scenarios 2.B.1 to 2.B.3, in 4 sample size points, 10, 20, 30, 100, in the 1st row. In the second row, is the average absolute Bias of the respective scenarios and the third row is the average of MSE. The gray lines in the first row reflect the true response rates. The prior distribution in this graph is having a mean of the 0.3} }
\label{fig:01_03_03prior}	
\end{figure}

    As described in the homogeneous scenarios, the choice of prior has an impact on the results, particularly by shifting estimates towards the prior mean. This effect is evident in fig \ref{fig:01_03_03prior}, which contrasts the results when using a prior mean of 0.3, compared to the prior mean of 0.5 shown in fig \ref{fig:01_03}. Most estimates shift accordingly to the homogeneous scenarios, illustrating the sensitivity of each method to the choice of prior.
    
    In general, the differences become more apparent with smaller cohort sample sizes, though no consistent ranking emerges among estimators. As sample sizes increase, estimators generally become more accurate and similar. Specifically, most methods converge to the true RR with cohort sizes of 100 or more. The advantages of these methods over the sample proportion in terms of lower MSE diminish in larger sample contexts.

    %Overall, Berry's BHM estimator appears more efficient in the small heterogeneity scenario (in terms of MSE), particularly with small sample sizes. However, as the heterogeneity increases, no clear winner emerges among the Bayesian estimators, with similar performance across estimators for reasonable sample sizes. The results are summarized in table \ref{tab:results}. 

A concise summary of these results can be found in Table 3.

\clearpage

\section{Discussion}   
    In this paper, we present a comprehensive review of Bayesian methodology for a basket-type design with single arm cohorts from the perspective of estimation. We evaluated various Bayesian methods and included the cohort sample proportion for reference.  We limited the evaluation to the setting of parallel single-stage cohorts using identical sample sizes. The main objective of this paper is to assess how these methods perform with respect to estimation of success proportions for the different cohorts in the basket trial. To ensure fair comparisons, we use flat or weakly informative prior distributions, following the original author's recommendations. The evaluation of the estimates considers frequentist properties including mean absolute bias and average mean squared error (MSE) and shrinkage metrics. The motivation for this research is the limited availability of research results that evaluate estimation properties when information borrowing is applied, whereas such estimation is particularly relevant for early phase basket trials that inform subsequent larger (confirmatory) trials.

    In a basket trial, statistical borrowing from other cohorts to improve estimation of the response rate for an individual cohort should essentially serve to improve the estimation compared to the cohort (sample) proportion. It is typically considered when the cohorts target rare conditions, and thus sample sizes per cohort are relatively small. The rationale of the basket trial (e.g. shared molecular target) typically provides justification for such borrowing. Even when such (mechanistic) justification is strong, heterogeneity in response rates between cohorts usually cannot be excluded a priori. Therefore, key desirable properties of Bayesian methods for estimation in this setting are twofold: improved precision at cohort level when heterogeneity is limited or absent, and sufficiently sensitive to adapt (i.e., limit bias at cohort level) when heterogeneity between cohorts is clearly present. 

    When put against these criteria, we conclude that there is no clear winner among the Bayesian methods in terms of optimal average MSE and average absolute bias across the scenarios evaluated. (See also table \ref{tab:results}) Berry's et al. \cite{Berry} demonstrates the smallest average MSE and average absolute bias when the true response rate belongs to the homogeneous scenario. When a higher heterogeneity level is introduced there is no optimal choice. Except for large cohort sample sizes, overall bias and mean squared errors are relatively substantial in some cases, especially as heterogeneity increases. The methods differ in the amount of shrinkage and in the amount that the estimate is influenced by the prior. All the methods shrink towards the overall mean across the cohorts. Berry's et al. \cite{Berry}, Jin's et al. \cite{Jin}, and Chen \& Lee \cite{Lee} estimates shrink more towards the overall mean compared to other Bayesian estimates. Fujikawa et al. \cite{Fuji} and Liu et al. \cite{Liu_kane} seem to be pulled towards 0.5 compared to the other estimates, as clearly seen in fig \ref{fig:Total_est_mean}. Chen \& Lee \cite{Lee} estimate, on the other hand, differs from the other estimators as it appears to shrink towards 0 if the overall true response rate mean is smaller than 0.5 and towards 1 if this is larger is over 0.5. Given the set criteria Berry's et al. \cite{Berry}, Fujikawa et al. \cite{Fuji} and Liu et al. \cite{Liu_kane} methods are less preferred, and EXNEX, Psioda's, Jin's et al. \cite{Jin}, and Chen \& Lee \cite{Lee} methods are more suitable, with the EXNEX method to be more consistent than the other methods. Notably, the performance of the sample proportion is equal or superior compared to the Bayesian methods in terms of MSE when the heterogeneity increases, and of course it is unbiased. In this paper, the focus was on evaluating the methods under a setting where there is no formal prior information concerning the homogeneity and the heterogeneity structure that could be included in the estimation. Using more informative priors does fit within some of methods, which could lead to more precise outcomes. 

    The complexity of the methods introduces a potential limitation to the present work and applications. The tuning parameters needed for the different methods imply that there is substantial flexibility (hence heterogeneity) in their implementation, without clear a priory guidance to optimally set these parameters. We used default values and non-informative priors in a simple setting, but in practice other choices may be made. The non-informative prior distributions has a prior mean of 0.5, when we made a different prior mean choice, we see a different behaviour in the estimated results. 
    
    In our effort to (computationally) replicate these methods for our simulation study, we faced many difficulties in selecting the appropriate parameters and ensuring that our implementation was indeed exactly equal to the published results and scripts and also occasionally found discrepancies between the publicly shared script and the scientific paper (that were subsequently resolved). Researchers may face similar struggles in determining the method most suitable for each situation and ensuring an appropriate computational implementation. Additionally, fully understanding the methods and their implementation is not straightforward, with usually guidance lacking in addressing estimation.

    Several simplifying assumptions were used in the simulation study. The choice for 6 cohorts of equal sample size with a single-stage design only covers a limited number of potential scenarios. In practical settings, it is common to suggest at least one interim analysis (e.g., following Simon's two-stage designs). In the single-stage setting, the sample proportion is an unbiased estimator, hence it was included as reference in our study. The proposed Bayesian methods do allow for interim analyses. Jin et al. \cite{Jin} proposed the use of a single interim analysis, while Fujikawa et al. \cite{Fuji} and Psioda et al. \cite{Psioda} allow several interim futility assessments. Berry et al. \cite{Berry} proposed an interim analysis when a certain number of patients are included (e.g., 10) and more assessments allowed after a pre-specified number of patients (e.g., 5). Simon et al. \cite{Simon_bay} considers the futility assessment after each observation. Neuenschwander et al. \cite{EXNEX} and Chen \& Lee \cite{Lee} do not propose an interim analysis stage, but an interim analysis could be applied at any time point of the trial. We did not evaluate the resulting very broad range of possible scenarios, which may lead to some differences between the methods when interim analyses are implemented. However, as the number of cohorts with similar response rates investigated correspond to realistic practical settings and a range of sample sizes was explored, we do believe the present provides basis for an initial choice of methods.

    The explored methods offer a potentially valuable tool for researchers in efficiently designing and analyzing basket trials. Estimators based on methods that allow borrowing of information across cohorts introduce bias, but are expected to have a smaller MSE, due to an increase in precision. However, there are many available options even within each method, with most methods requiring choices of tuning parameters in addition to priors for model parameters. A priory guidance on precise settings of these parameters for practical applications is challenging, which may limit the possibility to pre-specify the full estimation procedure. A simulation study, such as the one performed here, but targeted to a specific context of a particular study, could give a better insight.  

\clearpage

\section*{Appendix A}

\begin{landscape}
\begin{table}[ht]
\centering
\setlength{\tabcolsep}{5pt} % Adjust column separation
\renewcommand{\arraystretch}{1.2} % Adjust row separation
\begin{tabular}{p{3.2cm} p{3.5cm} p{5cm} p{5cm} p{4cm}}
\toprule
\textbf{Methods} & \textbf{Prior effect} & \textbf{Homogeneous scenario} & \textbf{Heterogeneous scenario} & \textbf{Shrinkage} \\
\midrule

Psioda & \raggedright bias toward the prior mean \arraybackslash & \raggedright Small MSE, slightly biased estimate \arraybackslash & \raggedright Borrowing to the mean and bias to the prior \arraybackslash & \raggedright Moderate to small shrinkage in all scenarios \arraybackslash\\

 Berry & \raggedright no effect \arraybackslash & \raggedright Smallest MSE, no bias \arraybackslash & \raggedright Bias estimate to the total mean \arraybackslash & \raggedright Extreme shrinkage to the total mean \arraybackslash\\

EXNEX & \raggedright limited - no influence \arraybackslash & \raggedright Small MSE \arraybackslash & \raggedright Shares less information when 1 vs 5 true RR basket, small MSE \arraybackslash & \raggedright Small shrinkage in small heterogeneity. Almost no shrinkage in high heterogeneity \arraybackslash\\

 Jin (CBHM) & \raggedright Slight influence \arraybackslash & \raggedright Small MSE, slightly biased \arraybackslash & \raggedright Slight prior influence, Small MSE \arraybackslash & \raggedright Moderate shrinkage in small heterogeneity. Small shrinkage in high heterogeneity \arraybackslash \\

Chen \& Lee (BCHM) & \raggedright No influence \arraybackslash & \raggedright Small MSE, almost unbiased, estimates to 0 if true RR $<$ 0.5 or towards 1 if true RR $>$ 0.5 \arraybackslash & \raggedright Shrinkage to the total mean, the smallest MSE \arraybackslash & \raggedright Small shrinkage in small heterogeneity. Moderate shrinkage in high heterogeneity \arraybackslash\\

Fujikawa & \raggedright bias toward the prior mean \arraybackslash & \raggedright Bias to prior mean, small MSE \arraybackslash & \raggedright Extreme bias to the prior mean \arraybackslash & \raggedright Moderate shrinkage in small heterogeneity. Small shrinkage in high heterogeneity \arraybackslash\\

Liu & \raggedright bias toward the prior mean \arraybackslash & \raggedright Bias to prior mean, moderate MSE \arraybackslash & \raggedright Extreme bias to the prior mean \arraybackslash & \raggedright Almost no shrinkage regardless the heterogeneity level \arraybackslash\\

Sample prop & \raggedright  \arraybackslash & \raggedright The largest MSE \arraybackslash & \raggedright MSE close to the Bayesian methods, in highly heterogeneous trials smaller MSE \arraybackslash & \raggedright No shrinkage \arraybackslash\\

\bottomrule
\end{tabular}
\caption{Results overview table}
\label{tab:results}
\end{table}
\end{landscape}

\clearpage
\section*{Appendix B}

\subsection*{Prior and parameters choice}
    The choice of the prior is a hard task to handle, especially since the choice usually affects the estimation. We made a choice to use weakly informative or uninformative priors to compare the methods under the same rules, following the suggestions of the respective authors for uninformative priors to use in their method. 

    Jin et al. \cite{Jin}, Berry et al. \cite{Berry}, and EXNEX \cite{EXNEX} do recommend somewhat informative priors involving pre-specified response rates $q_0$ and $q_1$ that correspond to inactive and active treatments respectively. Many of the methods are originally designed for a setting in which the goal of the study is to decide whether the treatment is active or not (represented by posterior probabilities of response rate $\geq q_1$ and $\leq q_0$ respectively). However, in our simulation study we focus on estimation of the response rates rather than decision making and use uninformative priors as specified below for each method.

\begin{itemize}

\item {The \textbf{\textit{Berry}} et al.\cite{Berry} method, proposes a strategy to provide non informative choices of hyper-parameters. For this case, we set the hyper-parameters as follows:
\begin{align*} 
\mu \sim N(\mu_0, \sigma_0^2) = N(0, 100) \\
\sigma^2 \sim IG(\lambda_1, \lambda_2) = IG(0.0005, 0.00005)
\end{align*}}

\item {The \textbf{\textit{EXNEX}}\cite{EXNEX} following the suggestions in the original paper we use a single exchangeability distribution of the EXNEX. The probabilities of each distribution is fixed, in the simple EXNEX model, the EX part is $p_i = 0.5$, and the NEX is $1-p_i = 0.5$. A normal prior distribution is proposed and the mean can reflect the expectation of the researcher in the logit scale. In order to make a similar choice with the methods that uses Beta(1,1) prior and the mean of this prior is the 0.5, we set the mean of the EXNEX prior to be 0, which is the analogous of 0.5 in the logit scale. 

\begin{align*}
\theta_i^{EX} \sim N(\mu_0, \sigma_0^2)\\
\mu_0 \sim N(0, 10) \\
\sigma_0^2 \sim half-normal(1) \\
(p_i , 1 - p_i) = (0.5,0.5)\\
\theta_i^{NEX} \sim N(m_i, v_i)\\
m_i = 0\\
v_i = 10
\end{align*}

}

\item {\textbf{\textit{Psioda's}}\cite{Psioda} method in the original paper is to set the a weakly informative Jeffreys prior $Beta(0.5,0.5)$. For the purpose of this paper, we use the uninformative $Beta(1,1)$. A prior model is set for the all possible models. The default settings of the package allow us to have a weakly informative prior where the model probabilities are giving greater weight to the more complicated models (less borrowing allowed). }

\item {\textbf{\textit{Fujikawa}}-like \cite{Fuji} method estimates uses a beta prior $Beta (a_j , b_j) = Beta(1,1)$. } and the amount of borrowing is tuned by the parameter of $\tau = 0.5$.

\item {\textbf{\textit{Jin}} et al.\cite{Jin} made an extensive simulation study to address the effect of weakly informative prior distributions on the proposed design, specifying each set of priors for the respective distance measure of the correlation matrix.

\textbf{Hellinger} (H) distance also introduces an exponential correlation, where the $\phi = Gamma(1.5,1)$ has a different prior mean. Similarly the parameters are specified by:
\begin{align*}
\theta_0 \sim N(\mu_0, \sigma_0^2) \\
\sigma^2 \sim IG(c_{\sigma^2}, d_{\sigma^2}) = IG(0.01,0.01) \\ 
\tau^2 \sim IG(c_{\tau^2}, d_{\tau^2}) = IG(0.01,0.01) \\ 
\sigma_0^2 \sim IG(c_{\sigma_0^2}, d_{\sigma_0^2}) = IG(0.1,0.1) 
\end{align*}

}

\item {\textbf{\textit{Chen \& Lee}} \cite{Lee} present the choice of the prior and the parameters used in their paper in detail. For the classification model a choice of non-informative conjugate normal distributed prior $\mu = 0.2$ and $\sigma_0^2 = 10$ is used to calculate the posterior probability of the true response rate. The parameter $\alpha = 10^{-60}$ and $\sigma_d^2 = 0.001$ choice can affect the cluster number, which are used in the Dirichlet process (DP). The hyper-prior parameters calculated in order to propose a non-informative prior choice. 

\begin{align*}
\theta_i \sim N( \mu_1, \frac{1}{\tau_1 m_i} ) \\
\mu_1 \sim N(\mu_2, \frac{1}{0.1}) \\
\tau_1 \sim Gamma(50, 10)
\end{align*}
}

\item {\textbf{\textit{Liu}} \cite{Liu_kane} proposed an uninformative prior for the posterior distribution which follows a $Beta(1,1)$. The level of leverage that each of the complicated models has, is chosen by the parameter $\delta$. We choose to use the $\delta = 0$, so each model is weighted equally.
}

\end{itemize}

\clearpage

\section*{Appendix C}

\clearpage

\section*{Appendix D}

\begin{figure}[h]
\centering
\includegraphics[width=17cm,height=15cm]{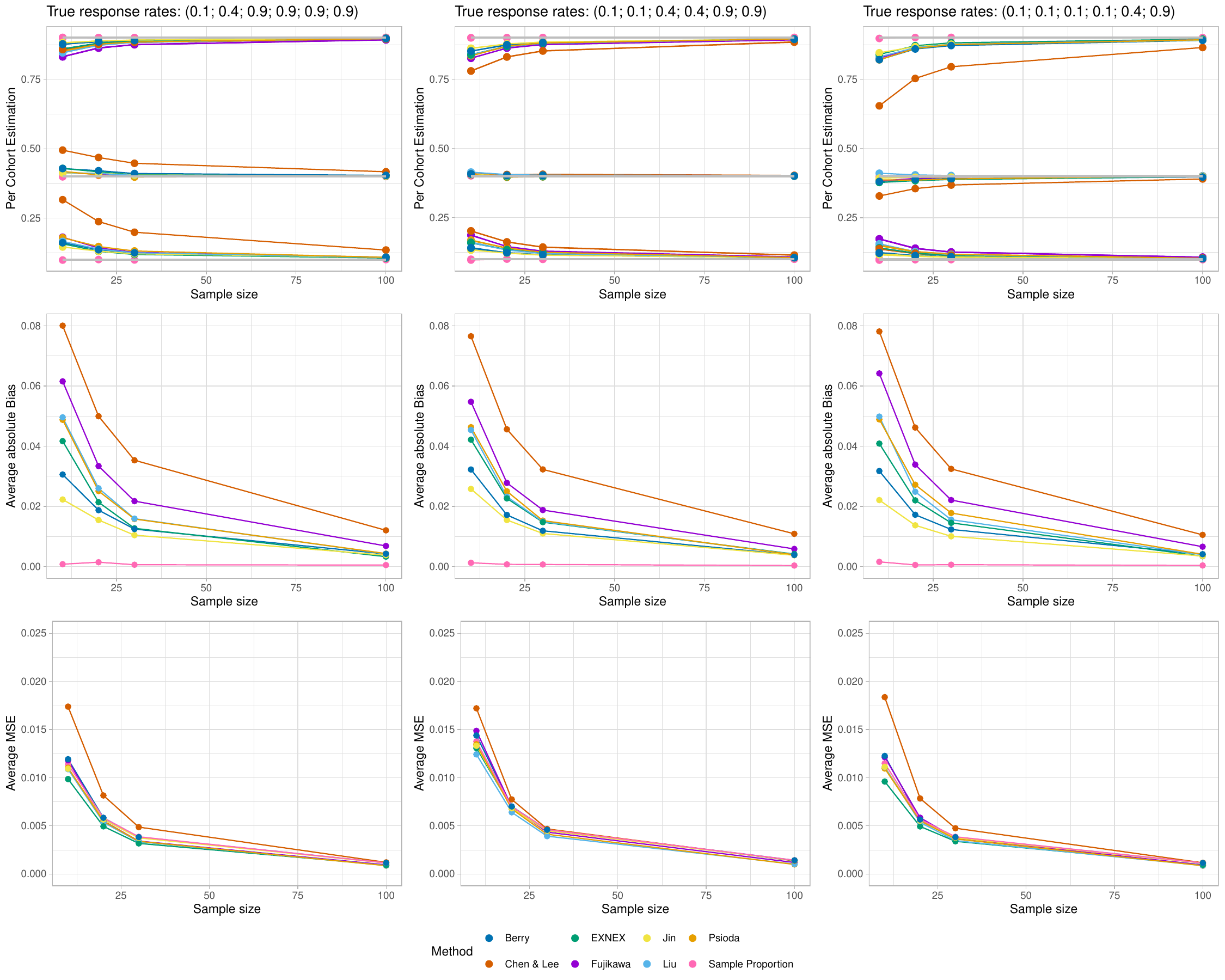}
\caption{\textit{Estimates on scenarios 3.B.1 to 3.B.3, in 4 sample size points, 10, 20, 30, 100, in the 1st row. In the second row, is the average absolute Bias of the respective scenarios and the third row is the average of MSE.} }
\label{fig:01_04_09}	
\end{figure}

\end{document}